\documentclass[twoside,twocolumn,9pt]{article}
\usepackage{extsizes}
\usepackage[super,sort&compress,comma]{natbib} 
\usepackage[version=3]{mhchem}
\usepackage[left=1.5cm, right=1.5cm, top=1.785cm, bottom=2.0cm]{geometry}
\usepackage{balance}
\usepackage{mathptmx}
\usepackage{sectsty}
\usepackage{graphicx} 
\usepackage{lastpage}
\usepackage[format=plain,justification=justified,singlelinecheck=false,font={stretch=1.125,small,sf},labelfont=bf,labelsep=space]{caption}
\usepackage{subcaption} 
\usepackage{float}
\usepackage{fancyhdr}
\usepackage{fnpos}
\usepackage[english]{babel}
\addto{\captionsenglish}{%
  
}
\usepackage{comment}
\usepackage{array}
\usepackage{droidsans}
\usepackage{charter}
\usepackage[T1]{fontenc}
\usepackage[usenames,dvipsnames]{xcolor}
\usepackage{setspace}
\usepackage[compact]{titlesec}
\usepackage{hyperref}

\usepackage{epstopdf}

\definecolor{cream}{RGB}{222,217,201}

\usepackage{amsmath,amssymb}
\usepackage{calrsfs}
\DeclareMathAlphabet{\pazocal}{OMS}{zplm}{m}{n}
\SetMathAlphabet\pazocal{bold}{OMS}{zplm}{bx}{n}

\begin{document}

\pagestyle{fancy}
\thispagestyle{plain}
\fancypagestyle{plain}{
\renewcommand{\headrulewidth}{0pt}
}

\makeFNbottom
\makeatletter
\renewcommand\LARGE{\@setfontsize\LARGE{15pt}{17}}
\renewcommand\Large{\@setfontsize\Large{12pt}{14}}
\renewcommand\large{\@setfontsize\large{10pt}{12}}
\renewcommand\footnotesize{\@setfontsize\footnotesize{7pt}{10}}
\makeatother

\renewcommand{\thefootnote}{\fnsymbol{footnote}}
\renewcommand\footnoterule{\vspace*{1pt}%
\color{cream}\hrule width 3.5in height 0.4pt \color{black}\vspace*{5pt}} 
\setcounter{secnumdepth}{5}

\makeatletter 
\renewcommand\@biblabel[1]{#1}            
\renewcommand\@makefntext[1]%
{\noindent\makebox[0pt][r]{\@thefnmark\,}#1}
\makeatother 
\renewcommand{\figurename}{\small{Fig.}~}
\sectionfont{\sffamily\Large}
\subsectionfont{\normalsize}
\subsubsectionfont{\bf}
\setstretch{1.125} 
\setlength{\skip\footins}{0.8cm}
\setlength{\footnotesep}{0.25cm}
\setlength{\jot}{10pt}
\titlespacing*{\section}{0pt}{4pt}{4pt}
\titlespacing*{\subsection}{0pt}{15pt}{1pt}

\fancyfoot{}
\fancyfoot[LO,RE]{\vspace{-7.1pt}\includegraphics[height=9pt]{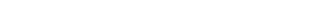}}
\fancyfoot[CO]{\vspace{-7.1pt}\hspace{13.2cm}\includegraphics{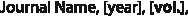}}
\fancyfoot[CE]{\vspace{-7.2pt}\hspace{-14.2cm}\includegraphics{RF.eps}}
\fancyfoot[RO]{\footnotesize{\sffamily{1--\pageref{LastPage} ~\textbar  \hspace{2pt}\thepage}}}
\fancyfoot[LE]{\footnotesize{\sffamily{\thepage~\textbar\hspace{3.45cm} 1--\pageref{LastPage}}}}
\fancyhead{}
\renewcommand{\headrulewidth}{0pt} 
\renewcommand{\footrulewidth}{0pt}
\setlength{\arrayrulewidth}{1pt}
\setlength{\columnsep}{6.5mm}
\setlength\bibsep{1pt}

\makeatletter 
\newlength{\figrulesep} 
\setlength{\figrulesep}{0.5\textfloatsep} 

\newcommand{\topfigrule}{\vspace*{-1pt}%
\noindent{\color{cream}\rule[-\figrulesep]{\columnwidth}{1.5pt}} }

\newcommand{\botfigrule}{\vspace*{-2pt}%
\noindent{\color{cream}\rule[\figrulesep]{\columnwidth}{1.5pt}} }

\newcommand{\dblfigrule}{\vspace*{-1pt}%
\noindent{\color{cream}\rule[-\figrulesep]{\textwidth}{1.5pt}} }

\makeatother

\twocolumn[
  \begin{@twocolumnfalse}
{\includegraphics[height=0pt]{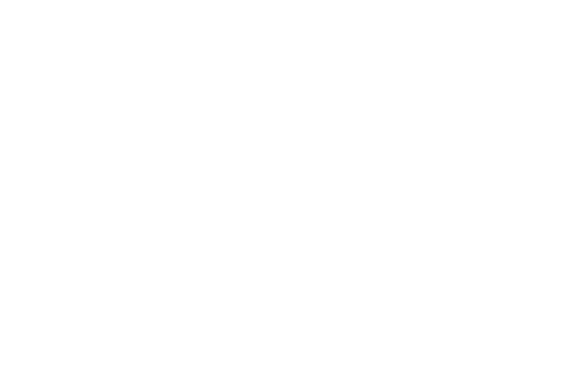}\hfill\raisebox{0pt}[0pt][0pt]{\includegraphics[height=0pt]{RSC_LOGO_CMYK.eps}}\\[1ex]
\includegraphics[width=18.5cm]{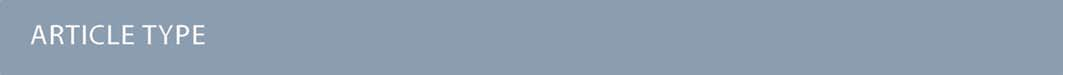}}
\par
\vspace{1em}
\sffamily
\begin{tabular}{m{4.5cm} p{13.5cm} }

\includegraphics{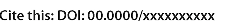} & \noindent\LARGE{\textbf{Collective motion of energy depot active disks}} 

\vspace{0.3cm} \\

\hfill & \noindent\large{Juan Pablo Miranda$^{a,b}$,Demian Levis$^{c,d}$,Chantal Valeriani$^{a,b}$ } \\

\includegraphics{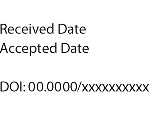} & \noindent\normalsize{%
In the present work we have studied collectives of  active disks with an energy depot, moving in the two-dimensional plane and interacting via excluded volume. The energy depot accounts for the extraction of energy taking place  at the level of each particle in order to perform self-propulsion, included in an underdampled Langevin dynamics. 
We show that this model undergoes a flocking transition, exhibiting some of the key features of the Vicsek model, namely, band formation
and giant number fluctuations. 
 Large density bands  disappear as the activity is further increased, eventually reaching a homogeneous polar state. 
We unravel an effective alignment interaction at the level of two-particle collisions that can be controlled by activity and gives rise to flocking at large scales. 
} 
\end{tabular}
\end{@twocolumnfalse} \vspace{0.6cm}
]

\renewcommand*\rmdefault{bch}\normalfont\upshape
\rmfamily
\section*{}
\vspace{-1cm}



\footnotetext{\textit{$^{a}$Departamento de Estructura de la Materia, Física Térmica y Electrónica, Universidad Complutense de Madrid, 28040 Madrid, Spain}}
\footnotetext{\textit{$^{b}$GISC - Grupo Interdisciplinar de Sistemas Complejos 28040 Madrid, Spain}}
\footnotetext{\textit{$^{c}$~Departament de F\'isica de la Mat\`eria Condensada, Universitat de Barcelona, Marti i Franqu\`es 1, 08028 Barcelona, Spain.}}
\footnotetext{\textit{$^{d}$~UBICS  University  of  Barcelona  Institute  of  Complex  Systems,  Mart\'{\i}  i  Franqu\`es  1,  E08028  Barcelona,  Spain.}}


\section{Introduction}
\label{sec:intro}

Active  matter systems, composed of a collection of interacting self-propelled units, have been the focus of a great deal of research efforts over the last  decades \cite{marchettiRev, shaebani2020computational,ComplexAndCrowdedEnvironments}. 
Given they are not in equilibrium, these systems  
exhibit a number of large-scale phenomena, not detected in 
equilibrium systems. A salient example is the emergence of collective motion, widely observed both in the living world and in synthetic realisations of active particles designed in the laboratory \cite{vicsek2012collective, chateDADAM}. Collective motion has been reported in a broad range of time and lenght scales, ranging from 
systems made of $10^{-7}-10^{-6}$m   objects (e.g. actomyosin networks \cite{schaller2010polar}, bacteria \cite{peruani2012collective, sokolov2012physical, li2019data} and colloidal suspensions \cite{bricard2013emergence, kaiser2017flocking}) up to systems composed of agents of the order of $\sim1$m in size (e.g. animal groups \cite{bialek2012statistical}). 

To gain a theoretical insight on this seemingly common feature displayed by a large variety of active systems of self-propelled units,  simple models have been proposed, such as Active Brownian Particles \cite{romanczuk2012active}, the Vicsek model \cite{VicsekOriginal} together with their continuum hydrodynamic  descriptions \cite{VicsekOriginal, tonerPRL95, ramaswamy2010mechanics, cates2015motility}. 
Spherical Active Brownian  particles have shown to exhibit phase separation even when  interacting only repulsively. This phase separation is induced by self-propulsion  and appears in dense enough situations  
\cite{cates2015motility}.  
Whereas the Vicsek model describes the behaviour of  self-propelled aligning particles, 
propelled at a constant speed  and solely interacting  via velocity-alignment interactions, being ferromagnetic/polar or nematic \cite{chateDADAM}. 
In two dimensions, a collection of 
polarly  aligning self-propelled particles exhibits a flocking transition towards a collectively moving state, the emergence of which is typically accompanied by large density heterogeneities in the form of travelling bands \cite{solon2015phase,Chat2008CollectiveMO}.   
Beyond the ideal case in which particles are just point-like, 
aligning particles have also demonstrated to undergo a  flocking transition, with the emergence of band-like patterns, even when  repulsive interactions are taken into account \cite{martinez2018, barre2015motility, sese2018velocity, sese2021phase}. These works suggest that  
neither excluded-volume interactions nor  collision rules are enough to change the phenomenology of the order-disorder transition
in  the Vicsek model.
Flocking can also arise from particles' shape (e. g. self-propelled rod-like particles aligning upon collision \cite{bar2020self, grossmann2020particle}) or from collisions between spherical particles which are not momentum conserving, such as  vibrated polar grains \cite{deseigne2010collective} (the latter are typically modelled as particles 'self' re-orienting 
their direction of self-propulsion with the one of their instantaneous velocity \cite{szabo2006phase, dauchot2019dynamics, dauchot2022collective, paoluzzi2024flocking}).

Although self-propulsion needs an energy intake from the environment (or from an internal fuel) to convert it into motion in the presence of dissipation,  Vicsek-type models and other simple active particles descriptions do not consider it explicitely. 
These models describe a system of active particles at a larger mesoscopic scale, somehow coarse-graining the microscopic details of the self-propulsion mechanism.  
One of the earliest model of active particles \cite{PRL98}  explicitly considering an internal energy intake   leading    to self-propulsion, is the so-called energy depot model 
\cite{schweitzer2003brownian}. 
The dynamics of such Energy Depot Active Particle (EDAP) model has been thoroughly studied in the past, mostly considering medium-mediated inter-particle interactions  \cite{lobaskin2013collective} or their dynamics in an external potential \cite{schweitzer2001statistical,mach2007modeling,ebeling1999active}. This model has proven useful for a 
stochastic thermodynamic description of active matter 
\cite{chaudhuri2014active, seifert2019stochastic}, even though 
most authors have used Active Brownian particles   \cite{szamel2019stochastic, shankar2018hidden} and Active Ornstein-Ulhenbeck Particles \cite{fodor2016far,caprini2019entropy} for this purpose.

As far as we know, the collective behaviour of dense suspensions of EDAP with excluded volume interactions has not been explored so far.    
Other works have focused on modifying the model considering different assumptions, leading to  substantial differences on diffusivity and transport properties, such as the effect of cross-correlated noise\cite{fang2021transport,guan2018transport}, 
a braking mechanism\cite{zhang2008active,zhang2009energydepotmembrane}, coupling a load to an energy depot particle \cite{ann2015stochastic} and even their motion  in disordered media where their propulsion is coupled to the properties of the latter \cite{olsen2021landscapes}.

\begin{figure}[h!]
\includegraphics[width=0.45\textwidth]{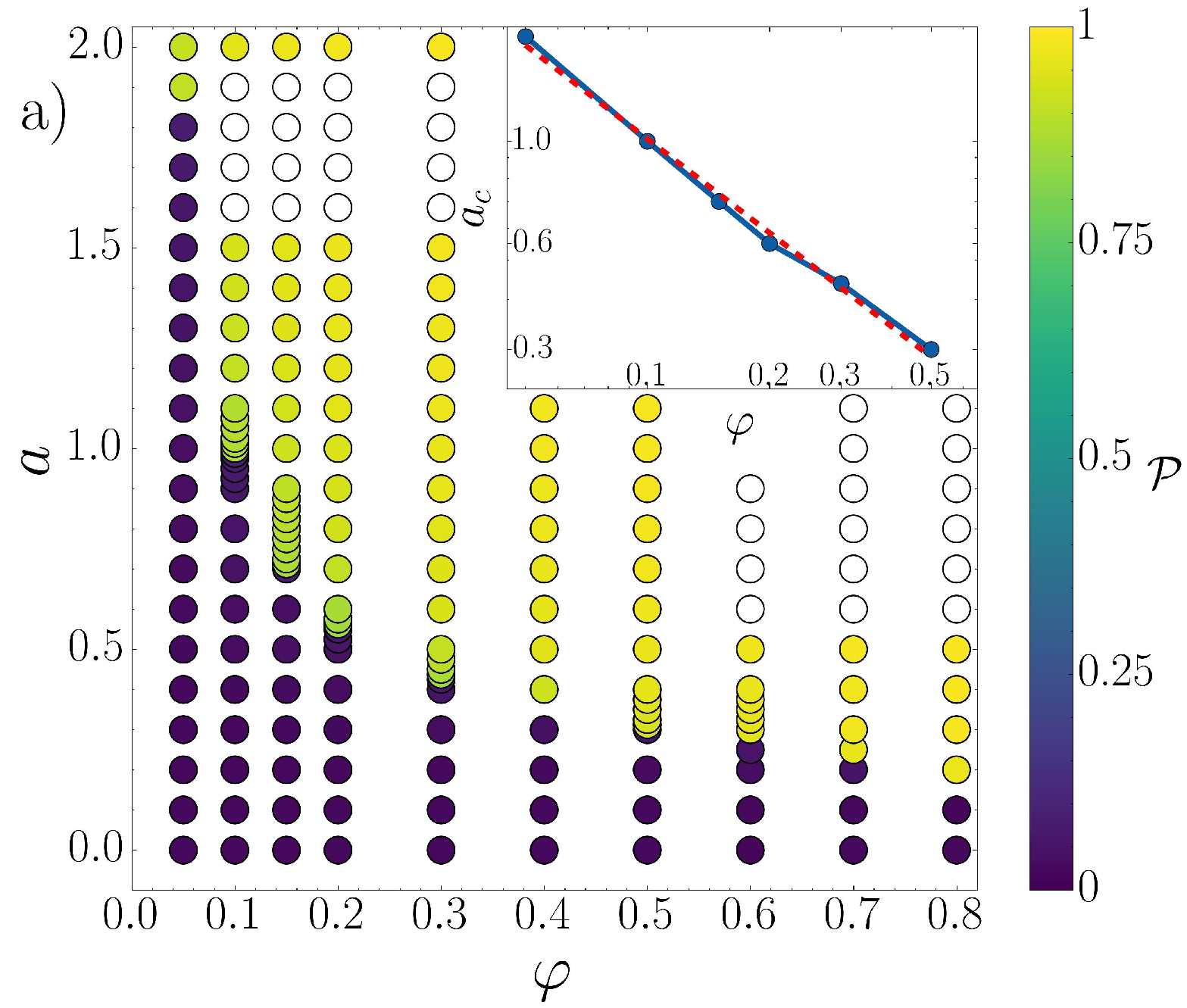}
\includegraphics[width=0.49\textwidth]{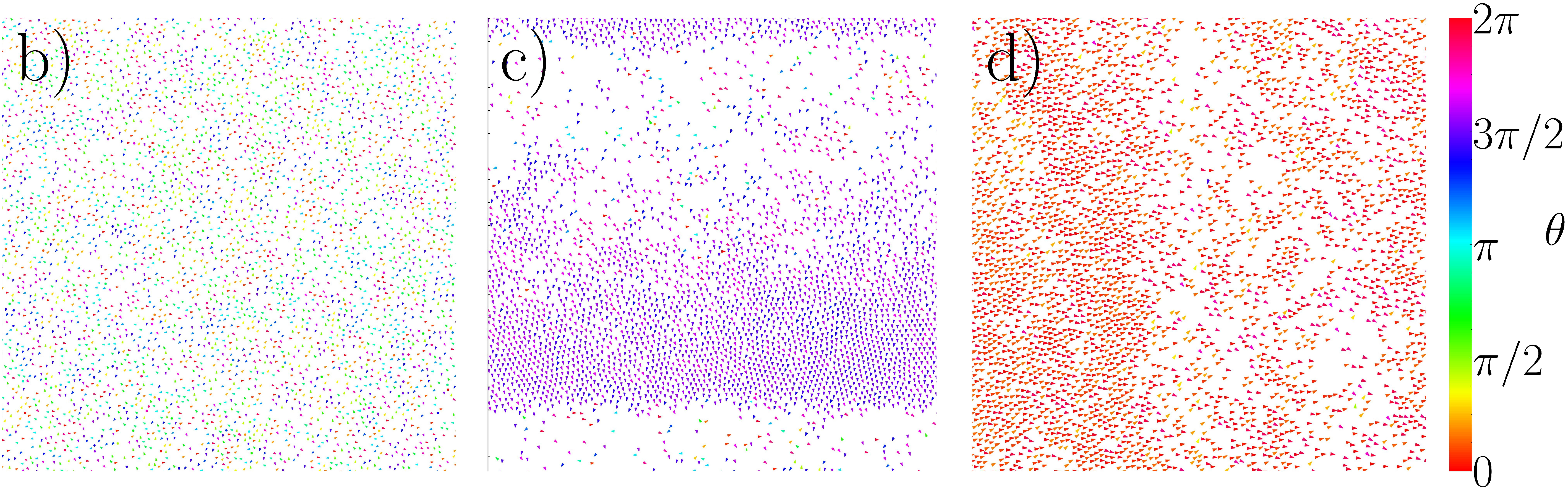}
    \caption{
    a) State diagram for a system of $N=2000$ particles representing at which activity and packing fraction the system is in a disordered or polar  state, as indicated by a color map of the steady global polarisation  (empty circles correspond to  parameters we did not explored). 
    The inset shows the  activity threshold $a_c(\varphi)$ (in log-log scale) above which we observe collective motion (extracted as the maximum of the susceptibility). The discontinuous red line corresponds to a power-law decay $\sim \varphi^{\lambda}$, with $\lambda=-0.77$.
    Snapshots of the steady state  of $N = 10000$ systems with $\varphi =0.3$: b) $a=$ 0.3, homogeneous and disordered  state ; c) $a=$ 0.5, a dense polar band surrounded by a (dilute) disordered background;  d) $a=$  0.9, polar state with a more homogeneous density distribution.} 
 \label{fig:statediagram}
\end{figure}

In the present work, we consider  repulsive spherical EDAP (disks)  in two-dimensions. We show that the mere competition between crowding effects and self-propulsion is enough to trigger a flocking transition, exhibiting  some of the key features of Vicsek model's phenomenology, namely band formation and giant number fluctuations. As shown in the 'density-activity' plane, 
 Fig. \ref{fig:statediagram}, 
beyond a  given threshold in density and/or activity,  the system sets in a collectively moving state. The transition towards such state is accompanied by the formation of large density bands that disappear as the activity is further increased, eventually reaching a homogeneous polar state. To understand this behaviour, we extract an effective interaction via the radial distribution function and find that 
 pair-collisions lead to an effective alignment  that can be controlled by  activity and  is responsible for the flocking transition to occur. 

The paper is structured as follows: in sec. \ref{sec:activeparticleswithanenergydepot} we define the model and its different regimes; in sec. \ref{sec:collectivemotion} we focus on the identification and characterisation of  collective motion; in sec.  \ref{sec:effectivealignent} we shed light on the origins of the reported flocking behaviour in the absence of explicit alignment, showing that an effective alignment interaction emerges from two-particle collisions. 

\section{Model for self-propelled particles with an energy depot}
\label{sec:activeparticleswithanenergydepot}
\begin{figure}[h!]
    \centering
    \includegraphics[width=0.45\textwidth]{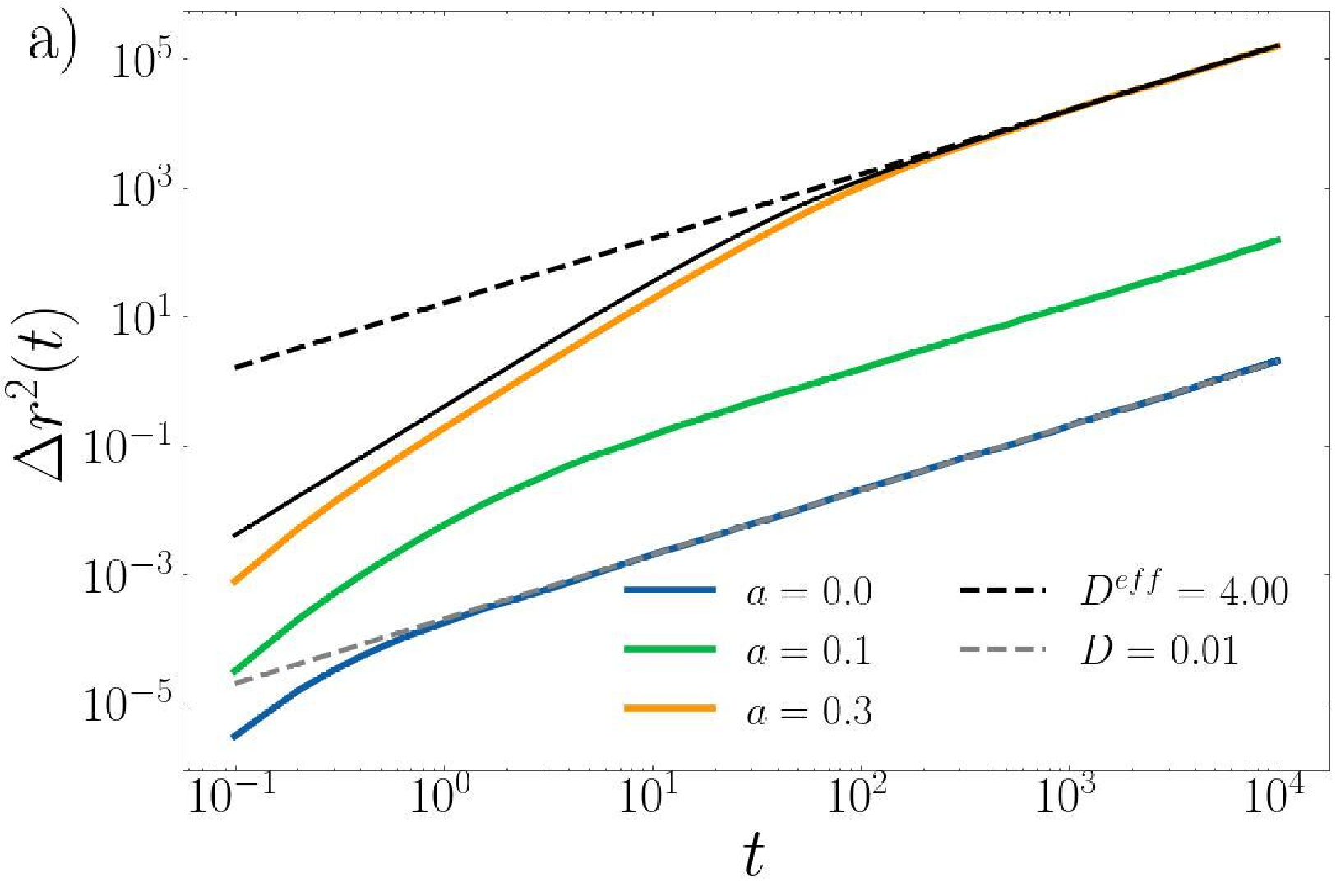}
    \includegraphics[width=0.45\textwidth]{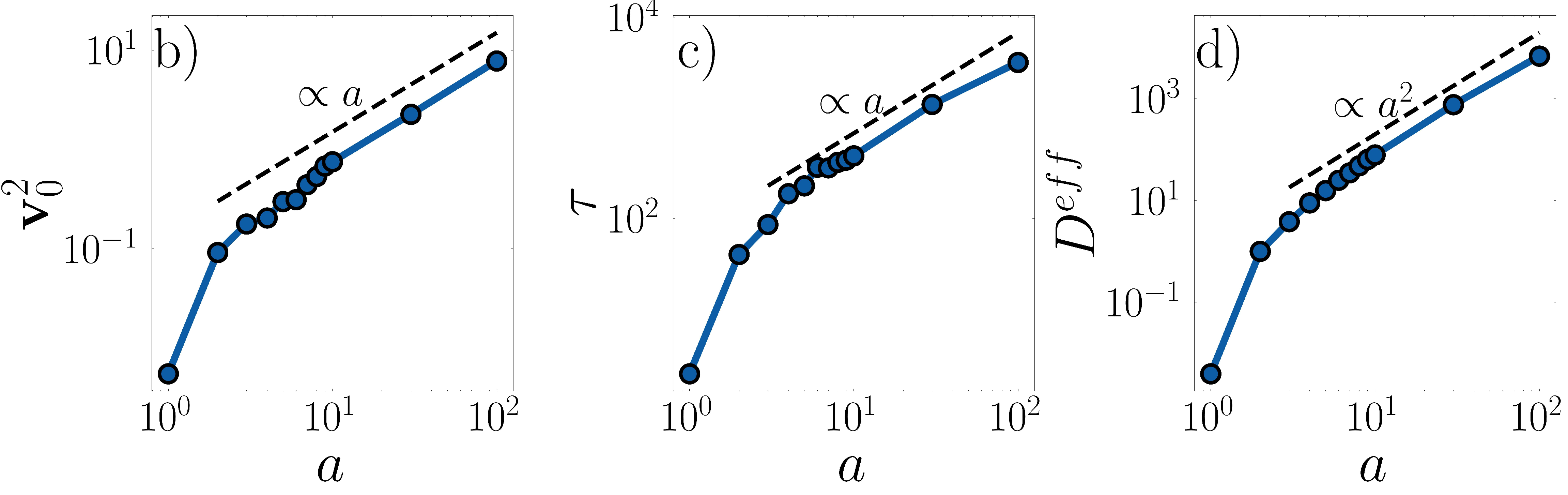} 
    
    \caption{a) MSD  of non-interacting energy depot particles for different degrees of activity at fixed $\gamma_0=10$, $k_BT=5.10^{-4}$ and $b=1/10$.  
    The black solid line represents the analytical expression \eqref{msd_depot} for a system with $a=0.3$. Discontinuous lines represent the long time diffusive behaviour with the effective diffusivity given in the key.  
    Blue line represents a system with $a=0$, i. e, in the Brownian case.  
    The green line corresponds to the limit case $a=b$. The red line represents a system in the  regime $a>b$ with $a=3$. 
 We extract  by fitting our data to eq. \ref{msd_depot}  $\mathbf{v}^2_0$ (b), $\tau$ (c) and $D^{\mathrm{eff}}$ (d), and compare them to the expected scaling with $a$ . }
\label{fig:msd_noninteracting}
\end{figure}

We consider a two dimensional suspension of 
 $N$ particles whose position and velocities are  $\mathbf{r}_i$ and  $\mathbf{v}_i$, respectively. 
Each particle carries an internal energy depot $\varepsilon_i$ , describing the conversion of energy extracted from the environment into kinetic energy at the level of individual particles. The dynamics of $\varepsilon_i$  is overdamped and given by \cite{PRL98,schweitzer2003brownian}
\begin{equation}
    \dot{\varepsilon}_i(t)=q-c\varepsilon_i(t)-d{\textbf{v}}_{i}^2\varepsilon_i(t)
\end{equation}
The parameters $q$, $c$ and $d$ quantify the rate of energy intake, energy dissipated and conversion into kinetic energy, respectively. The motion of each particle is then governed by the following Langevin equation 
\begin{equation}   \label{EDM}
  m \dot{\textbf{v}}_i=-(\gamma_0-d\varepsilon_i(t)) \textbf{v}_i - \sum_{j\neq i}\boldsymbol{\nabla} U(r_{ij}) + \sqrt{2D}\boldsymbol{\xi}_i(t)
\end{equation}
 where  $\boldsymbol{\nabla} U(r_{ij})$  is the inter-particle repulsive interaction and  $\boldsymbol{\xi}$  a Gaussian white noise of zero mean and unit variance, mimicking a thermal bath at temperature $T$. 
The damping coefficient $\gamma_0$  (obeying $\gamma_0D=k_BT$) combines to the internal energy  depot term. 
  The energy depot acts on each particle as  a self-propulsion   force: indeed, it appears as an advective term, which  can be cast as an effective velocity-dependent damping coefficient $\gamma(\mathbf{v}_i)=\gamma_0-d\varepsilon_i(t)$. If $\gamma(\mathbf{v})<0$  the particle is accelerated by the effect of the energy depot, while in the opposite case it is still damped but with a smaller damping coefficient than $\gamma_0$. 
In the adiabatic limit, considering that the energy depot is the fastest degree of freedom,  the equations of motion simplify to 
\begin{equation}
  m \dot{\mathbf{v}}_i=-\gamma(\mathbf{v}_i) \mathbf{v}_i- \sum_{j\neq i}\boldsymbol{\nabla} U(r_{ij}) +\sqrt{2 D} \xi_i(t)
    \label{Langevin}.
\end{equation}
with
\begin{equation}
\gamma(\mathbf{v})  = \gamma_{0} \left( 1 - \frac{a}{ b + \mathbf{v}^2 } \right)
\label{gammav gamma0}
\end{equation}
where we have introduced,  to reduce the number of parameters,  $a = q /\gamma_{0}$ and $b = c/d$. 
EDAP exhibit different dynamic regimes depending on the values of these parameters \cite{schweitzer2003brownian}. 
We denote $\mathbf{v}_0=\sqrt{a-b}$ , the velocity for which $\gamma(\mathbf{v})=0$, with $a>b$. We can then identify two different situations depending on the value of the velocity with respect to this reference. For $\mathbf{v}>\mathbf{v}_0$ the motion is damped with  $0<\gamma(\mathbf{v}) < \gamma_0$ , approaching the passive limit as $\mathbf{v}$ increases. Meanwhile, for  $\mathbf{v}<\mathbf{v}_0$, $\gamma(\mathbf{v}) < 0$, meaning that the energy depot accelerates the motion in the direction of $\mathbf{v}$. Eventually, as the particle accelerates, it reaches a velocity  close to $\mathbf{v}_0$, above which its motion is damped. All in all, the average velocity of a single EDAP is $\mathbf{v}_0$ \cite{schweitzer2003brownian}.
For $a<b$, particles always exhibit damped Brownian motion as $\gamma(\mathbf{v})$  remains positive. 
The passive limit is  recovered when $a\to 0$, allowing for as smooth connection with a well-know equilibrium system. We shall thus identify $a$ as our \emph{activity} parameter, and consider the other energy depot parameters as fixed. 
In the non-interacting limit 
the mean-square displacement reads, in the weak noise (strong pumping) approximation  \cite{mikhailov_self-motion_1997,schweitzer2003brownian}
\begin{equation}
\Delta r^2 (t) = \langle(\mathbf{r}(t)-\mathbf{r}(0))^2\rangle \approx 4 D^{\mathrm{eff}} \left\{ t +  \tau \left[\exp{\left(-\frac{t}{\tau}\right)} - 1  \right]\right\}\, ,
\label{msd_depot}
\end{equation}
showing thus persistent random motion with effective diffusivity 
$D^{\mathrm{eff}} = \mathbf{v}_0^4 / 2k_B T \gamma_0 \propto a^2$
and persistence time 
$\tau = \mathbf{v}_0^2 / 2k_B T \gamma_0 \propto a $.
Thus, varying $a$ allows us to disentangle the role played by activity and go across the different dynamic regimes. 
In Fig \ref{fig:msd_noninteracting} we put eq.  \ref{msd_depot} into test, and run   numerical simulations to extract estimates of the parameters  $\mathbf{v}_0$, $\tau$ and $D_{\mathrm{eff}}$ (see \ref{fig:msd_noninteracting} panel (b)-(d)). 

The interactions between  particles derive from a  pairwise potential $U(r_{ij}=|{\textbf{r}}_i-{\textbf{r}}_j|)$, that here we choose to be purely repulsive and short-range. In practice, we use the following WCA form \cite{WCA}.
\begin{equation}
   U(\mathbf{r}_{i j})=\left\{\begin{array}{l}
4 u_0\left[\left( \frac{\sigma}{r_{i j}} \right)^{12}-\left(\frac{\sigma}{r_{i j}}\right)^{6}+\frac{1}{4}\right], \quad r_{i j} \leqslant 2^{1 / 6} \sigma \\
0, \quad r_{i j} \geqslant 2^{1 / 6} \sigma
\end{array}\right.
\label{eq:lennardjones}
\end{equation}
We simulate a system of $N$  active particles in an {$L\times L$} box  with periodic boundary conditions and  area fraction $\varphi=\frac{N \, \pi \,\sigma^2}{4 \, L^2}$. The Langevin dynamics has been implemented by means of  the  LAMMPS \cite{LAMMPS} open source package, making use of the Velocity Verlet integrator.
All physical quantities are expressed in Lennard-Jones reduced units, with lengths, times and energies given in  terms of $\sigma=\tau =\epsilon=1$, where $\tau=\sqrt{m\sigma^2/\epsilon}$ and $m=1$. 
We have run simulations of systems with  $N = 2 \times 10^3$ up to $2 \times 10^4$ particles at different surface fractions $\varphi$, in the range $\varphi=0.05 \,...,\,0.8$.
In our simulations, we  set $\gamma_0 = 10$, the time step to  
$\Delta t/\tau = 10^{-3}$  
and the reduced temperature    $k_B T/\epsilon= 5\times10^{-4}$. 
We explore the collective behaviour of the model at fixed $b=1/10$ as a function of both $a$ and $\varphi$. In practice, due to the extent of the parameter space in the model, we  choose to fix  $c$ , $d$ and $\gamma_0$, while the activity $a$ is varied  changing the values of $q$. As we group the parameters,
this choice is equivalent to varying the parameter $a$ while fixing $b$. 
The values of the original parameters were $c=1$,  $d=10$ and $q$ varying between $0$ to 30. 
This is equivalent to vary $a$ from $0$ to $3$. It has to be taken into account the fact that  systems with $a<b$ cannot be considered as active, even though  we  have also studied them.
The initial states have been prepared by setting particles and velocities at random, chosen from an uniform distribution.

\section{Transition to a flocking state}
\label{sec:collectivemotion}

To quantify  collective motion, we employ the global polar order parameter $\pazocal{P}$,  or polarisation, given by  the modulus of the average direction of the instantaneous velocity vector of each particle
\begin{equation}
\pazocal{P}
= \left\langle 
\left| \frac{1}{N} \sum_{i=1}^{N} \mathbf{e}_i  \right| \right\rangle ,
\label{eq:polar_order_parameter}
\end{equation}
where $\mathbf{e}_i$ is the orientation of the velocity vector of particle $i$,  given by: $\mathbf{e}_i = \mathbf{v}_i / \left| \mathbf{v}_i \right|=(\cos\theta_i,\sin\theta_i)$, and 
$\left\langle \cdot\right\rangle$ denotes a steady-state average.
Values of $\pazocal{P}>0$  indicate that a fraction of the particles is aligned, thus moving coherently, whereas $\pazocal{P}=0$ corresponds to particles moving in random directions.

\begin{figure}[h!]
\includegraphics[width=0.45\textwidth]{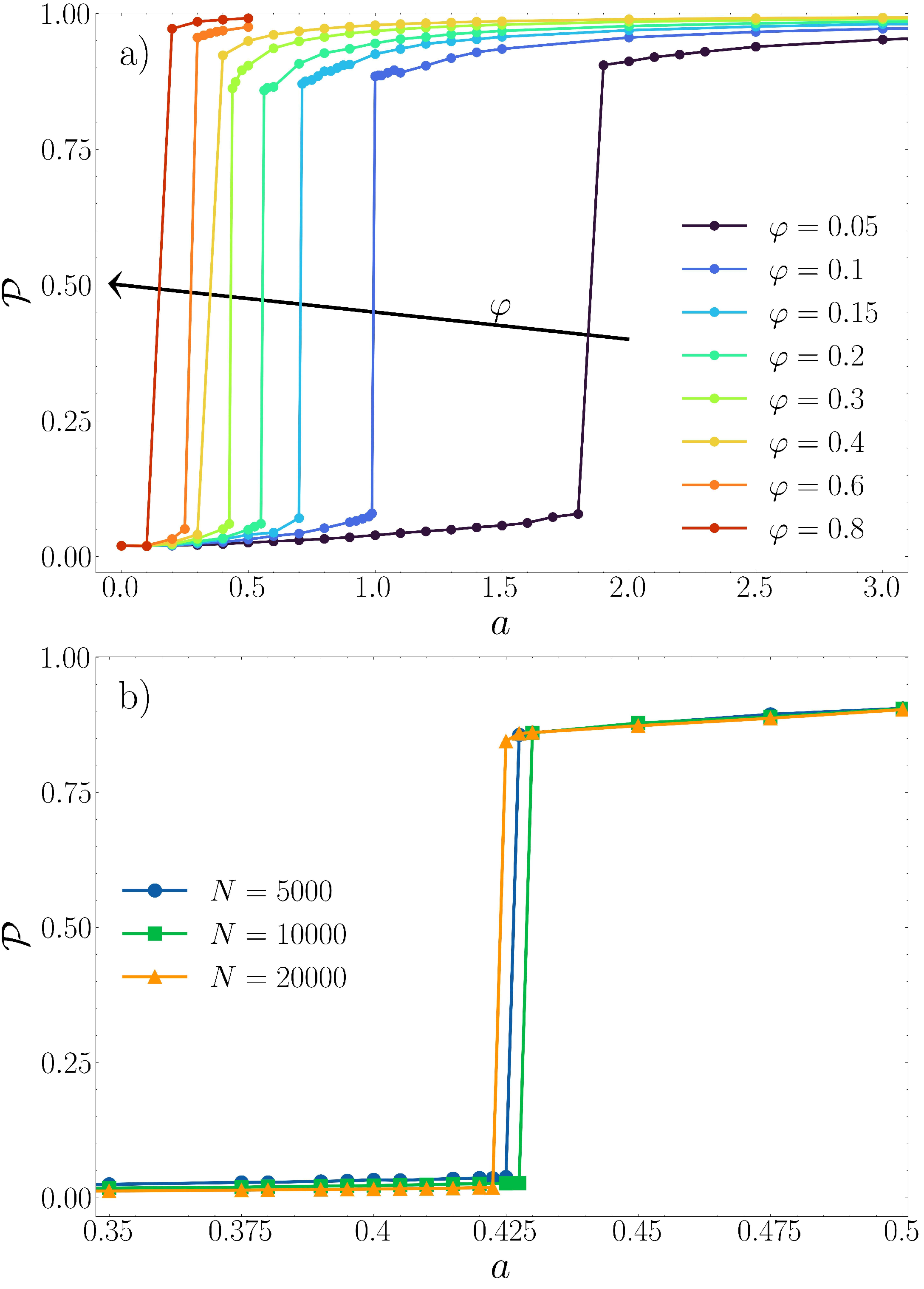}
\caption{\label{fig:Polar order interacting} 
  a) Polar order parameter as a function of $a$ for different $\varphi$ in a system of $N=2000$ particles. The different colors represent different area fractions $\varphi$ (the arrow indicates increasing values of $\varphi$).  b) Polarisation as a function of $a$ at fixed $\varphi=0.3$ for systems with $N= 5000$, $ 10000$ and $20000$. 
}
\end{figure}

Figure \ref{fig:Polar order interacting}-a) represents the values of the polarization  for a wide range of activities and packing fractions. 
Our main finding is the emergence of a polar state in  suspensions of  EDAP repulsive particles. To establish the robustness of our results, we have performed simulations at different system sizes. As shown in Figure \ref{fig:Polar order interacting}-b, we have studied the disorder-to-order transition in systems with   $N=5000,10000,20000$ while keeping the density fixed  ($\varphi = 0.3$), showing that the value of  activity needed for the transition to take place is not significantly affected by the system size in this parameter range. 
Interestingly, the polar order increases  with $a$ and $\varphi$, even though the system under study is not characterised by an explicit alignment akin  Vicsek-like models \cite{VicsekOriginal}.
More  examples of particles with a  polar order without an  explicit alignment have been reported earlier  \cite{CapriniLowenAtractiveABP_PRL23,choques,Lam_2015,dauchot2022collective}. 

For all densities reported in Fig.\ref{fig:Polar order interacting}-a), we detect a transition from a low value of   $\pazocal{P}$ at low activity $a$  (corresponding to a disordered state) to a $\pazocal{P} \rightarrow 1$ for larger  activity  $a$ (corresponding to a flocking state). The emergence of an alignment will be explained in details in section \ref{sec:effectivealignent} in terms of inter-particle collisions. Briefly, we have found that collisions favour alignment via the combined effect of self-propulsion in the direction of the particles' velocities and steric repulsion. When two particles collide, repulsive forces accelerate them longitudinally. Their speed is thus reduced and activity thus pushes them along a direction that will typically reduce the angle between their velocities.  As the rate of collisions increases with density, the onset of flocking decays with $\varphi$.

\begin{figure}[h!]
    \centering
    \includegraphics[width=0.5\textwidth]{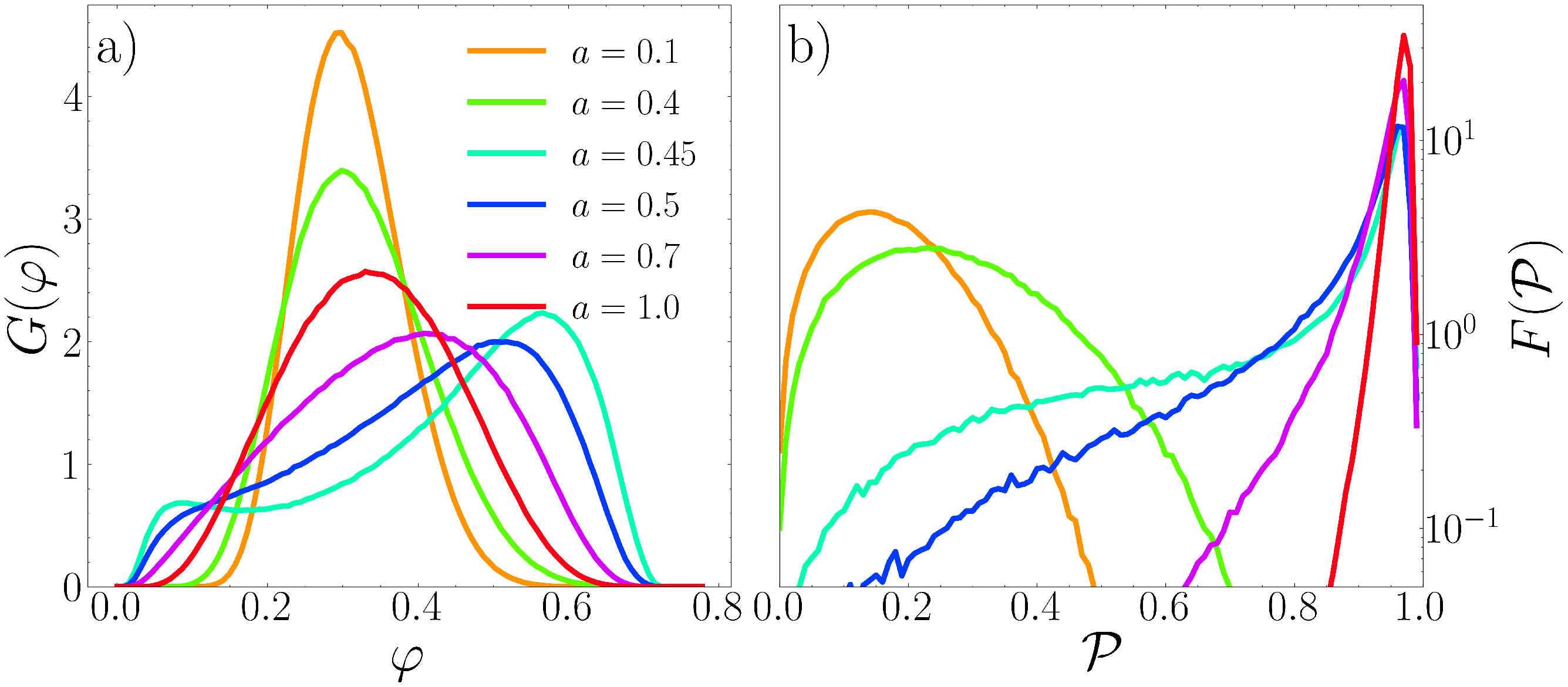}
    \caption{a) Local area fraction probability distribution $G(\varphi)$ calculated via Voronoi tessellations for systems of $N =10000$,  $\varphi = 0.3$ and different values of $a = 0.1..., 1$. b) Probability distribution function of the order parameter $F(\pazocal{P})$  (same parameters as panel a)).
 }
    \label{fig:PDFvoro}
\end{figure}

We show in  Fig.\ref{fig:statediagram}-a the region in parameter space where flocking occurs  (in yellow). The inset represents the activity threshold $a_c(\varphi)$ above which we detect collective motion:   $a_c$ decays with packing fraction (in a way that phenomenologically suggests $a_c\sim \varphi^{-0.77}$).  
To better understand the nature of the disorder-to-order transition, we characterise the structural properties of the different states. 
As shown in Fig.\ref{fig:statediagram}-b) the system displays different kind of structures in the steady state. These can be characterised using 
 the probability density of the order parameter $F(\pazocal{P})$ and the local packing fraction $G(\varphi)$.
$F(\pazocal{P})$ is obtained from the statistics of the local order parameter. In order to study the probability distribution of the packing fraction $G(\varphi)$, a Voronoi tessellation is performed to compute the local surface fraction of each configuration. 
 Based on the results of local density and  polarization, presented in Fig \ref{fig:PDFvoro}, we identify three states, reported in Fig.\ref{fig:statediagram}-b), Fig.\ref{fig:statediagram}-c), Fig.\ref{fig:statediagram}-d). 
The first one (Fig.\ref{fig:statediagram}-b) is the disordered state, which corresponds to a homogeneous density distribution  centered around the mean density (orange and green curves in Fig.\ref{fig:PDFvoro}-a), and a homogeneous local polarization distribution (orange and green curves in Fig.\ref{fig:PDFvoro}-b)  corresponding to the absence of  polar ordering. 
Increasing activity we detect a transition to an ordered phase. 
Beyond the onset of flocking one can identify two different states based on their local density and polarization. A heterogeneous state (Fig.\ref{fig:statediagram}-c) is observed close to the transition, for which   $G(\varphi)$  is no longer uni-modal, but displays two maximum values (one at low and one at high densities), signature of a dense region  coexisting with a dilute disordered background (see $a=0.45$ Fig.\ref{fig:PDFvoro}-a).
Interestingly, a heterogeneity in the distribution of polarisation coincides with this density heterogeneity. As shown in Fig.\ref{fig:PDFvoro}-b,  $F(\pazocal{P})$ exhibits a sharp peak close to 1 with a very broad tail at smaller values, signalling large fluctuations. Such states, as illustrated in Fig.\ref{fig:statediagram}-c) correspond to travelling bands, as typically observed in models of flocking. Bands are dense and very strongly polarised structures.

Increasing activity even further,  density heterogeneities disappear, and $G(\varphi)$ presents a uniform distribution again, although broader than in the small activity limit (see Fig.\ref{fig:PDFvoro}-a).
This regime corresponds to the emergence of a  homogeneous polar state (Fig.\ref{fig:statediagram}-d). The polarization distribution now exhibits a sharp peak close to 1, whose width decreases with increasing activity. 
So far, we have only detected band formation  for intermediate densities close to the transition,  quickly disappearing upon increasing activity.

As known from the literature of polar fluids \cite{polaractiveliquids}, giant number fluctuations typically appear in the flocking phase.
We measure the number density fluctuations to unravel possible connections between orientational order and giant number fluctuations. To do so, we divide the system into $2^k$  cells and count the number of particles in each cell, considering  a large number of configurations, to estimate 
the mean number of particles 
$\langle N \rangle$ and its variance $\Delta N$. We repeat the procedure over boxes with different sizes (ranging from from $k=2,\dots,8$) and  obtain a value of  $\langle N \rangle$ for each box size. We then extract a power-law relation  $\Delta N \sim N^{\alpha}$ where  a value $\alpha>1/2$ signals  giant number fluctuations.

\begin{figure}[h!]
    \centering
    \includegraphics[width=0.45\textwidth]{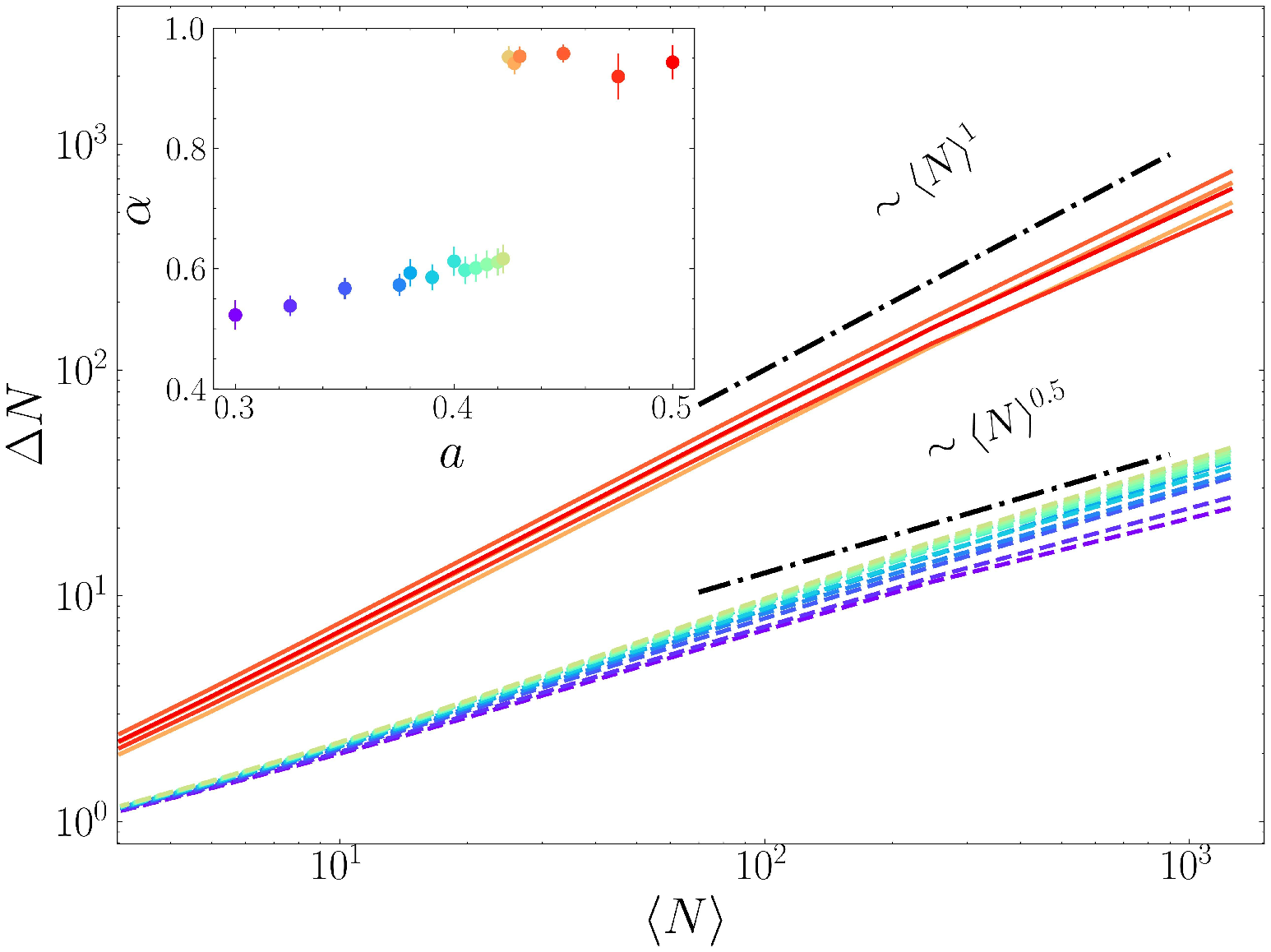}
    \caption{Variance of the Number of particle $\Delta N$ compared to the mean Number of particles $\langle N \rangle$ for systems of $N=20000$, $\varphi=0.3$ and different values of $a$ ranging from 0.3 to 0.5. Dashed lines represent the systems below the transition with values for $a$ range from $0.3$ to $0.4225$, while solid lines represent the systems that show polar order with corresponding values of $a$ from $0.425$ to $0.5$. The inset shows the dependency of  the $\alpha$ exponent with  $a$ as extracted from the data  in the main panel. }
    \label{fig:number-fluctuations}
\end{figure}

Figure \ref{fig:number-fluctuations} shows the variance of the number of particles $\Delta N$ as a function of the mean number of particles for systems at fixed density ($\phi=0.3$) but varying the value of the activity parameter. In all cases, $\Delta N \sim N^{\alpha}$. However, as clearly shown in the inset, depending on the activity, the exponent $\alpha$ noticeably varies, jumping from values around $1/2$ to almost 1 when $a\approx a_c$.  When $\alpha > 0.5$ the system exhibits giant number fluctuations. As shown in Fig.\ref{fig:number-fluctuations}, this occurs beyond the onset of flocking. 
When $\alpha \approx 0.5$, number fluctuations follow what one expects from the central limit theorem. The variation on the value of $\alpha$ with the activity parameter thus provides a complementary way of locating the onset of flocking.

\section{Understanding the effective alignment}
\label{sec:effectivealignent}

Even though many body interactions might be relevant to understand  collective phenomena in active particle systems  \cite{turci2021phase}, one might start with unravelling the emergence of collective motion from the analysis of  two body interactions. 

\begin{figure}[h!]
     \centering
         \includegraphics[width=0.22\textwidth]{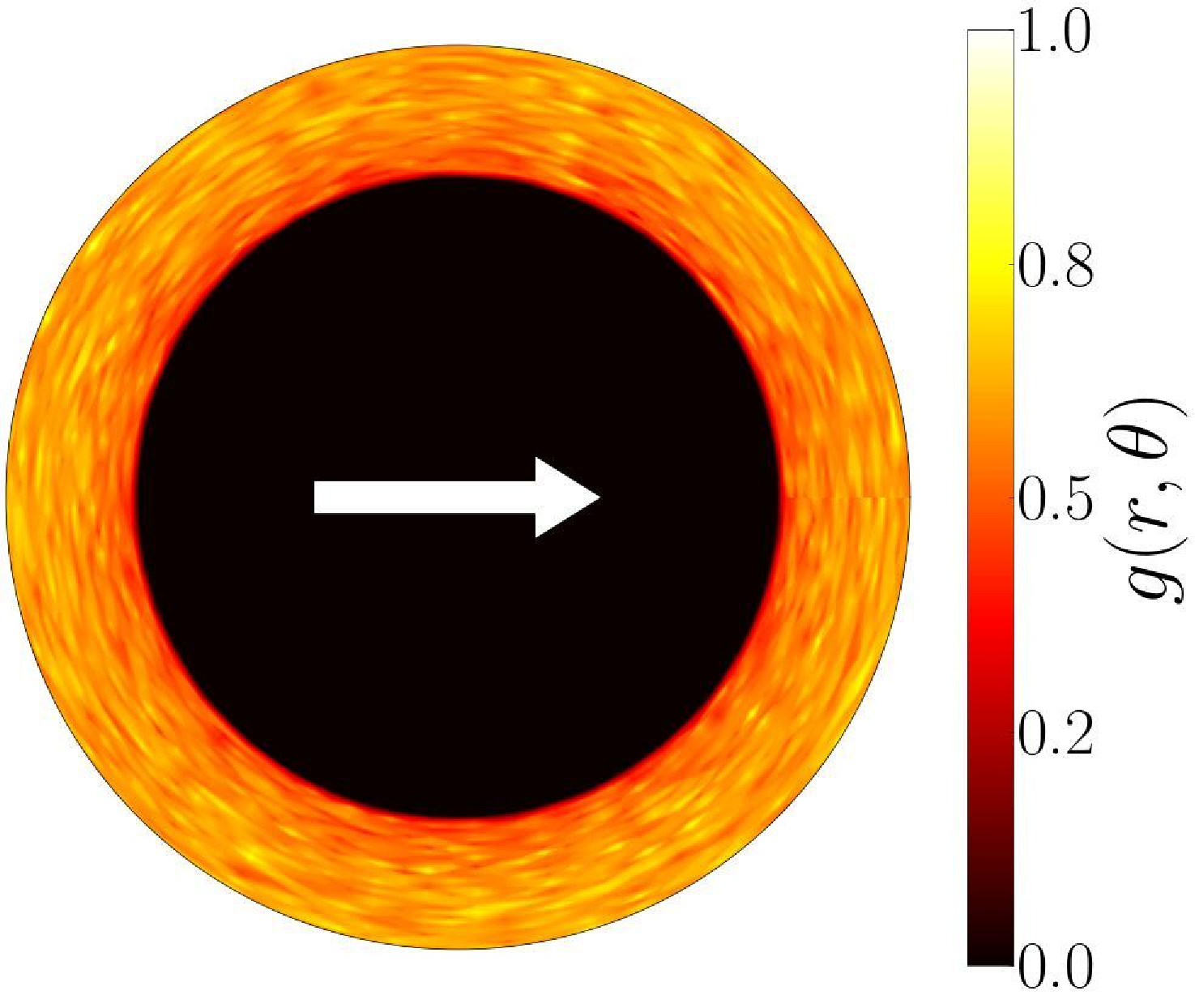}
         \includegraphics[width=0.22\textwidth]{angular_radial_distribution_q10.eps}
        \caption{Pair distribution $g(r,\theta)$ function for  $a = 0$ (left) and $a =1$ (right). The angles represent relative orientation between velocities and the color plot the numerical values. The thick arrow indicates the perfectly align direction. }
        \label{fig:angular radial} 
\end{figure}

To understand whether  the emergent behaviour is due to pair interactions, we  compute the two particle pair correlation function, depending on both the interparticle distances and the relative orientation of their velocities. 
This function can be defined  as \cite{hansen2013theory,bialke2013microscopic}
\begin{equation}
g({r},\theta) = \frac{V}{N^2} \left\langle \sum_i \sum_{j \neq i} \delta({r} - {r}_{ij}) \delta(\theta - \theta_{ij}) \right\rangle
    \label{rdf_definition}
\end{equation}
where ${r}_{ij}=|\mathbf{r}_{i}-\mathbf{r}_{j}|$ is the distance between two disks centres and $\theta_{ij}=\theta_{i}-\theta_{j}$ is the relative angle formed by their velocities.
We numerically computed $g({r},\theta)$ from two particle simulations.  The results are shown in  Fig. \ref{fig:angular radial}. 
While in the passive system (left-hand side) $g({r},\theta)$  is isotropic, in the presence of activity  (right-hand side) it shows a strong anisotropy, with a pronounced peak for values close to $\theta=0$, fading  to zero as the relative angle moves away. This means that particles' velocities are more likely to point along the same direction.  
Such correlation is localized at contact, and quickly decays for $r$ larger than the particle's radius. This shows that the competition between self-propulsion and steric effects gives rise to an effective alignment. 

\begin{figure}[h!]
    \centering
    \includegraphics[width=0.45\textwidth]{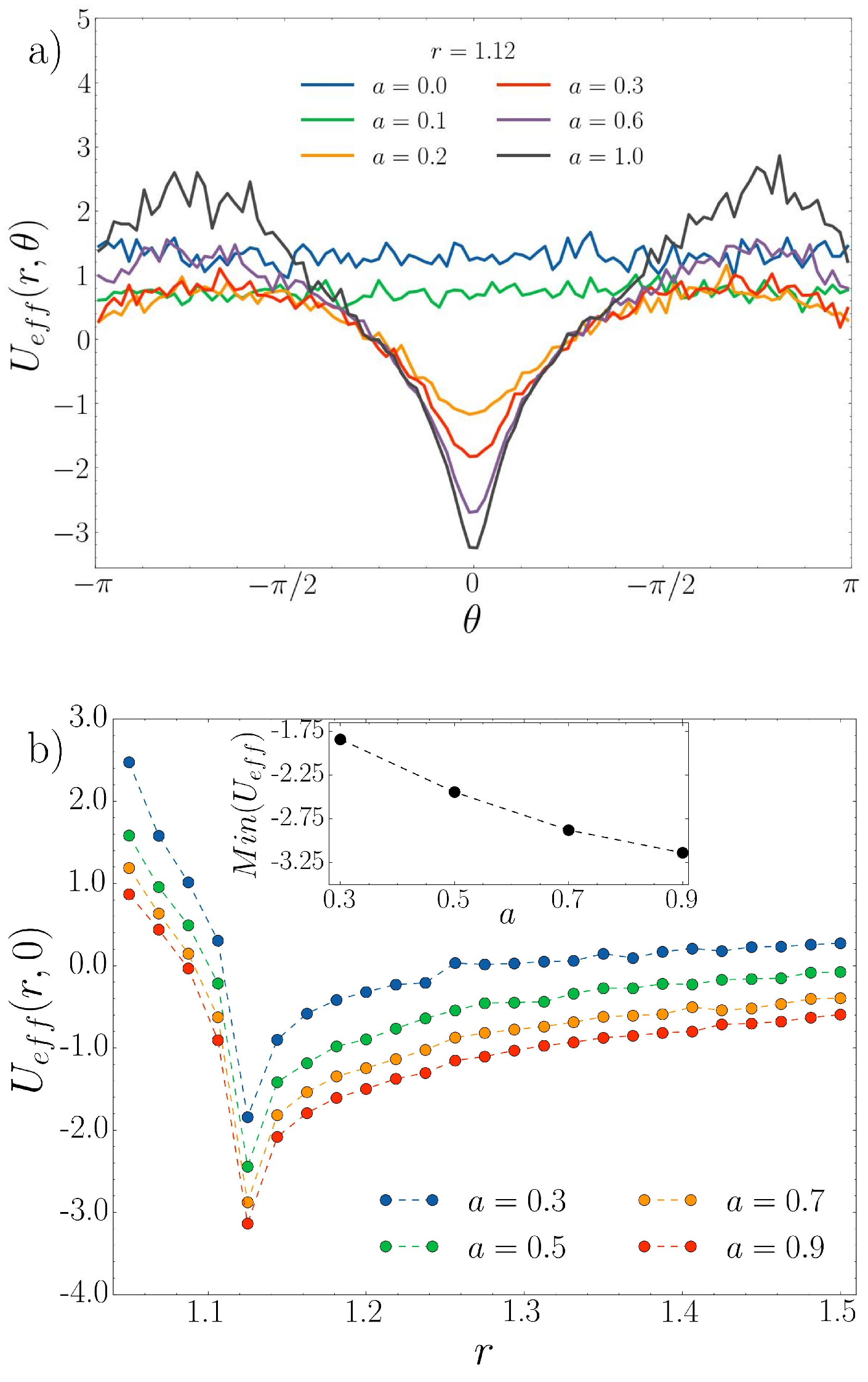}
    \caption{a): Effective potential as a function of the relative orientation between particle's velocities at a given interparticle distance $r=1.12\sigma$ and different values of $a$. b): Effective two particle potential as a function of distance for $\theta=0$ and different values of $a$. The inset on the figure shows how the minimum of the potential becomes more pronounced  with activity.}
    \label{fig:effective potential r and theta0}
\end{figure}

To further quantify such alignment, we extract an effective potential from the radial distribution \cite{farage2015effective,marconi2016effective} and study its behaviour in terms of the different parameters (such as the activity).
Assuming that $g(r,\theta) = \exp \left(- \beta U_{\mathrm{eff}}(r,\theta) \right)$,  one can  map the radial distribution function to an effective potential. Fig \ref{fig:effective potential r and theta0}. 
shows the effective potential as a function of the angle (panel a) and the effective potential as a function of the inter-particle distance, for varying values of the activity parameter $a$ (panel b). 
On the one hand, as reported in Fig \ref{fig:effective potential r and theta0} a), activity lowers the minima of the potential, favoring an effective alignment: $U_{\mathrm{eff}}$ is lowered as $a$ increases for relative angles close to 0, thus increasing the effective alignment as  self propulsion is increased.
 On the other hand, as shown in Fig \ref{fig:effective potential r and theta0} b), the minimum of  the effective potential corresponds to  the cutoff of the WCA potential $r = 2^{1/6}\sigma \simeq 1.12 \sigma$.
This  reveals that  alignment arises from the interplay between the excluded volume  interactions and activity.

Since the only interactions are given through excluded volume, the mechanism for alignment has to be mediated by contact interactions (collisions), explaining why the minima of the effective potential occurs at the contact distance between two particles. 
This also has to do with the fact that the other part of the mechanism is the active force, as the effective potential develops a deeper minimum as activity is increased.
Upon collision, a  particle $i$ will exert a force to another particle $j$  as 
\begin{equation}
     \mathbf{F}_{ij}^{\mathrm{col}} = -\gamma \left(  \mathbf{v}_{i} \right) \left(\mathbf{v}_{i} \cdot \hat{\mathbf{r}}_{ij} \right) \hat{\mathbf{r}}_{ij} - \boldsymbol{\nabla} U(r_{ij})
\end{equation}
The term $\gamma \left(  \mathbf{v}_{i} \right) \left(\mathbf{v}_{i} \cdot \hat{\mathbf{r}}_{ij} \right) \hat{\mathbf{r}}_{ij}$ corresponds to the projection of  $  -\gamma(\mathbf{v}_i) \mathbf{v}_i$ on the unit vector $\hat{\mathbf{r}}_{ij}=(\mathbf{r}_{i} - \mathbf{r}_{j})/|\mathbf{r}_{i} - \mathbf{r}_{j}|$. 
The other contribution corresponds to the force exerted by the WCA potential.
 Note here that $\gamma \left(  \mathbf{v}_{i} \right)$ is negative, 
 while the potential will create a force along the opposite direction. 

\begin{figure}
    \centering
    \includegraphics[width=0.49\textwidth]{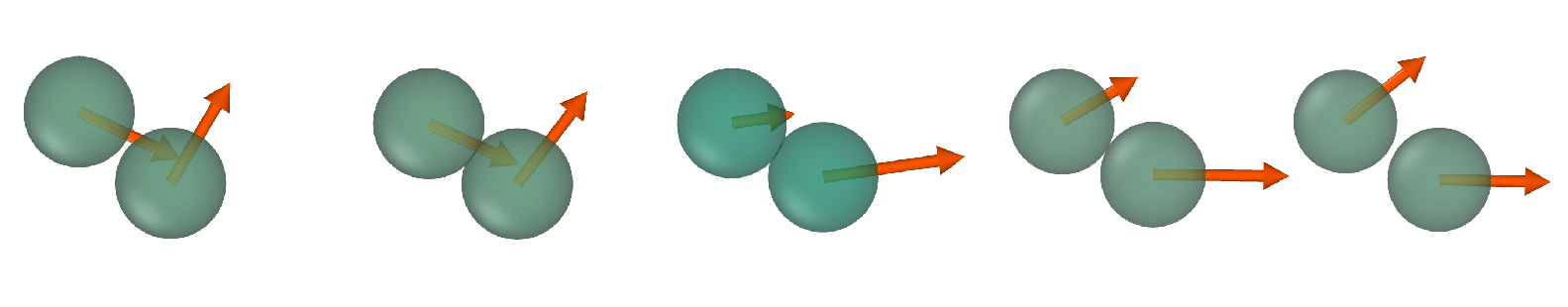}
    \caption{Successive snapshots during a collision between two particles, where their velocities are represented as orange arrows. The snapshots are ordered from left to right, as  time goes on.}
    \label{fig:colision2particulas}
\end{figure}

As this system is formed by self propelling disks, when they move in different directions with intersecting trajectories, they exert a torque on each other due to $\mathbf{F}_{ij}^{\mathrm{col}}$. The total torque acting on particle $j$ is thus 
\begin{equation}
\mathbf{M}_j = \mathbf{r}_{ij} \times  (\mathbf{F}_{ij}^{\mathrm{col}}- \gamma(\mathbf{v}_j) \mathbf{v}_j)   
\end{equation}
As shown in Fig. \ref{fig:colision2particulas},  the mutual torque  rotates particles' velocities,  pushing them towards the direction orthogonal to $\hat{\mathbf{r}}_{ij}$ (the torque vanishes in this case). As a result,  particles' velocities align along the same direction. Such simple mechanical argument sheds light into the effective potential $U_{\mathrm{eff}}$ extracted earlier: its minimum is located around contact  and becomes more pronounced as the term $\gamma(\mathbf{v}_j) \mathbf{v}_j$  in the torque becomes larger.  
To illustrate such mechanism we show in Fig. \ref{fig:colision2particulas} a sequence of snapshots during a collision event. The top-right particle  hits the other one from behind and it is slowed down by the interaction potential . This makes  the collision lasting longer than if the two particles were passive, since  the active force is pushing the particle to collide again.

\section{Conclusion}
The present work studied a system of active  disks self-propelled by an internal energy depot, characterised by a velocity dependent friction and an isotropic repulsive (hard) potential.
The novelty of this version of the EDAP model with respect to others is that it presents polar alignment, despite  repulsive interactions and not an explicit alignment mechanism.
A Vicsek-like transition from a non-polar isotropic state to a polar and ordered state is  identified and characterised in terms of the  polar order parameter and the giant number fluctuations. The  structural analysis reveals the presence of different phases, such as a non-polar homogeneous density phase, and two polar phases. The polar phases are divided  in a homogeneously dense and a band phase.

Even though the behaviour resembles that observed    in a suspension of self propelled Vicsek particles\cite{Chat2008CollectiveMO}, differences are that while in the Vicsek system the order parameter is controlled by the noise, in our system  ordering is related to  activity, promoting the polar order transition.
Given that the alignment mechanism is not explicit, an explanation is provided in terms of an effective potential originated by collisions between particles. The effective potential reveals a connection between the potential and the activity, which is translated  as a torque acting on particles.

From our results, we conclude that this model is a good candidate 
for more realistic systems of active engines\cite{fodor2021active} at the micro-scale, since it allows a theoretical approach that could complement numerical simulations (as in \cite{sekimoto2010stochastic,sekimoto1998langevin,Chaudhuri2013StochasticABP}).



\section*{Acknowledgements}

C.V. acknowledges fundings  IHRC22/00002 and 
PID2022-140407NB-C21 from AEI/MICINN.
D.L. acknowledges AEI/MICINN for funding under Project PID2022-140407NB-C22.
J.P.M thanks Miguel Barriuso and José Martín-Roca for helpful scientific discussions.
\section*{Data Availability Statement}
The data that support the findings of this study are available from the corresponding author upon reasonable request.

\balance

\bibliographystyle{rsc} 

\providecommand{\noopsort}[1]{}\providecommand{\singleletter}[1]{#1}%
\providecommand*{\mcitethebibliography}{\thebibliography}
\csname @ifundefined\endcsname{endmcitethebibliography}
{\let\endmcitethebibliography\endthebibliography}{}

\end{document}